\documentclass[aps,twocolumn,floatfix,superscriptaddress,nofootinbib]{revtex4-2}
\usepackage{booktabs}
\usepackage{graphicx}
\usepackage{braket}
\usepackage{stix}
\usepackage{xcolor}
\usepackage{appendix}
\usepackage{comment}
\usepackage{amsmath}
\usepackage{dcolumn}
\usepackage{soul}
\usepackage{bm}
\usepackage{amsthm}
\usepackage{enumitem}

\DeclareMathOperator*{\argmin}{arg\,min}
\usepackage{xr-hyper}
\usepackage{hyperref}
\hypersetup{colorlinks,
	citecolor=blue
}
\usepackage{cleveref}
\usepackage{apptools}
\AtAppendix{\counterwithin{lemma}{section}}

\makeatletter
\newcommand*{\addFileDependency}[1]{
  \typeout{(#1)}
  \@addtofilelist{#1}
  \IfFileExists{#1}{}{\typeout{No file #1.}}
}
\makeatother
\newcommand*{\myexternaldocument}[1]{%
    \externaldocument{#1}%
    \addFileDependency{#1.tex}%
    \addFileDependency{#1.aux}%
}

\myexternaldocument{Supp_Info}

\begin{document}

\preprint{APS/123-QED}

\title{Magnetic Phases of Spatially-Modulated Spin-1 Chains in Rydberg Excitons: Classical and Quantum Simulations}

\author{Manas Sajjan}
\affiliation{Department of Chemistry, Purdue University, West Lafayette, IN 47907}

\author{Hadiseh Alaeian}
\email{halaeian@purdue.edu}
\affiliation{Elmore Family School of Electrical and Computer Engineering, Purdue University, West Lafayette, IN 47907}
\affiliation{Department of Physics and Astronomy, Purdue University, West Lafayette, IN 47907}

\author{Sabre Kais}
\email{kais@purdue.edu}
\affiliation{Department of Chemistry, Purdue University, West Lafayette, IN 47907}
\affiliation{Department of Physics and Astronomy, Purdue University, West Lafayette, IN 47907}
\affiliation{Elmore Family School of Electrical and Computer Engineering, Purdue University, West Lafayette, IN 47907}

\date{\today}

\begin{abstract}
In this work, we study the magnetic phases of a spatially-modulated chain of spin-1 Rydberg excitons. Using the Density Matrix Renormalization Group (DMRG) technique we study various magnetic and topologically nontrivial phases using both single-particle properties like local magnetization and quantum entropy as well as many-body ones like pair-wise N\'eel and long-range string correlations. In particular, we investigate the emergence and robustness of Haldane phase, a topological phase of anti-ferromagnetic spin-1 chains. Further, we devise a hybrid quantum algorithm employing Restricted Boltzmann Machine to simulate the ground state of such a system  which shows very good agreement with the results of exact diagonalization (ED) and DMRG.
\end{abstract}

\maketitle


\section{Introduction}
Quantum spin systems have been the subject of intense studies for a long time as they offer a playground for exploring fascinating physics~\cite{Parkinson10}. 
In 1983, Haldane showed that chains of half-integer and integer spins are profoundly different~\cite{Haldane1983,HALDANE1983464}. In particular, while the Heisenberg antiferromagnetic chains have a ground state with an excitation gap and exponentially decaying correlations for integer spins, the ground state of half-integer spin chains has no excitation gap and the correlation decays algebraically. Since then, the rich physics of spin chains have been investigated extensively leading to the discovery of intriguing features such as half-integer spin edge modes~\cite{Hagiwara1990,Glarum1991} and non-local string orders~\cite{Nijs1989,Kennedy1992}.

Within the last two decades, the significant advances in the control of various quantum systems have led to the emergence of quantum simulators plethora in the well-controlled laboratory setups based on ultracold atoms in optical lattices, superconducting circuits, and quantum dots, to name a few~\cite{Lewenstein2007, Georgescu2014}. Such controllable quantum systems have been successfully used for the study of spin systems properties such as magnetism~\cite{Struck2011,Simon2011,Greif2013, Brown2015}, transport~\cite{Hild2014}, and topology~\cite{Duca2015,Jotzu2014,Xu2018, Sompet2022}. 
Recently, there have been several attempts to simulate various condensed-matter models using an array of Rydberg atoms, i.e. highly-excited atoms with macroscopic sizes and very susceptible to the external fields~\cite{Gallagher2005}. The enhanced polarizability of Rydberg atoms results in large, long-range interactions and makes them a suitable platform for the investigation of various spin systems~\cite{Weimer2010,Labuhn2016,Scholl2021,Ebadi2021}.

On the other hand, excitons, i.e. the quasi-particles of bounded electron-hole pairs in semiconductors, offer an alternative platform for the study of many-body quantum systems in a low-dimensional integrable and scalable platform at high densities not accessible with ultracold atoms. 
Like atoms, unique favorable scaling such as strong long-range interaction and Rydberg blockade effect are expected for Rydberg excitons, as well. The first observation of Rydberg excitons up to $n = 25$ in cuprous oxide (Cu$_2$O) in 2014~\cite{Kazimierczuk2014} revived the idea of excitonic quantum simulators and the field of semiconductors Rydberg physics~\cite{Assmann2020}.  

Here, we study the emergent magnetic phases of a generalized spin-1 chain. The Hamiltonian parameters and the coupling coefficients are based on values attainable with Rydberg excitons in Cu$_2$O, where the spin-1 degrees of freedom can be mimicked with optically-active $p$-excitons. Furthermore, we develop a hybrid quantum-classical algorithm to map this generalized spin-1 Hamiltonian to an spin-$\frac{1}{2}$ Ising hamiltonian using Restricted Boltzmann Machine (RBM) architecture and simulate the ground state of some prototypical examples.

The paper is organized as follows. In Sec.~\ref{sub: model} we introduce the model describing the most general dynamics of a spin-1 chain with nearest-neighbor interactions in a spatially-modulated potential. In Sec.~\ref{sub: DMRG} we employ the density matrix renormalization group (DMRG) to study various magnetic phases emerging in this system with physical parameters based off Rydberg excitons in Cu$_2$O. We classify the aforementioned many-body phases using pair-wise and long-range string correlations highlighting the emergence of N\'eel order and topological Haldane phase at different modulated potentials. 
Sec.~\ref{sub: RBM_sim} puts forward an a hybrid algorithm for the efficient simulation of studying stationary states on a NISQ hardware. Since we are studying spin-1 particles the Hilbert space grows as $3^N$ making the problem intractable rather quickly. We propose an algorithm that makes this scaling linear in the system size $N$ which allows us to simulate the exact dynamics of the model on a quantum computer. 
%
We summarize the results and conclude the work in Sec.~\ref{sec: conclusion} and discuss the follow-up directions. 

The results of this work pave the way toward the study of spin-1 chains, supporting diverse topological and magnetic phases, on an analog excitonic quantum simulator. Further, it puts forward an efficient algorithm to study the stationary states of such systems on a gate-based digital quantum computer for the first time.
\section{The model}~\label{sub: model}
In this section, we introduce the model of spin-1 chains characterized by the number of spins, $N$. The most general Hamiltonian with nearest-neighbor interaction is as follows~\cite{Klumper1993}
\begin{eqnarray}
    H &=&  \sum_{j = 0}^{j_{max}} c_0 + c_1 S_j^z S_{j+1}^z + c_2 (S_{j}^xS_{j+1}^x + S_{j}^y S_{j+1}^y)\nonumber\\
    &+& c_3 (S_{j}^z S_{j+1}^z)^2 + c_4 (S_j^xS_{j+1}^x \nonumber\\
    &+& S_j^y S_{j+1}^y)^2 + c_5(S_j^z S_{j+1}^z (S_j^x S_{j+1}^x \nonumber\\ &+& S_j^y S_{j+1}^y) + H.c.) \nonumber \\ &+& c_6 (S_j^x S_{j+1}^x - S_j^y S_{j+1}^y)^2 \,,\label{eq: Ham_haldane}
\end{eqnarray}
where $S_i^\alpha$ are the components of the usual S = 1 spin operators with $\alpha \in \{x,y,z\}$. For the periodic boundary conditions (PBC) , $j_{max} = (N-1)$ with $j_{max}+1 = 0$ whereas for the open boundary conditions (OBC), $j_{max}=(N-2)$. Several well-known spin models like the transverse field Ising model \cite{PhysRevB.42.2597, PhysRevB.67.172402}, commonly-studied Heisenberg spin models like XXZ \cite{Gomez1989,LIU201499} or XYZ\cite{albayrak2022antiferromagnetic,ALBAYRAK20111631}, and the AKLT  model with bi-quadratic interaction terms \cite{Affleck1987,Solyom1987,Klumper1992,Batchelor1990,Fannes1992} are special cases of Eq.~\ref{eq: Ham_haldane} with specific choices of coefficients $\{c_i\}_{i=0}^6$.

The realization of such a spin model on an array of Rydberg excitons in cuprous oxide has been proposed recently~\cite{Poddubny2019, Poddubny2020} with the promise of attaining topologically non-trivial phases. Such excitons are created through optical excitations into Rydberg states of varying principal quantum numbers $n$ with the azimuthal quantum number $l=1$ ($p$-shell). The latter endows the array to behave like an effective pseudospin-1 particle with the van der Waals interaction between excitons leading to the terms in Eq.~\ref{eq: Ham_haldane}, under the assumption that only one exciton occupies each trapping site and the direct exchange interaction is negligible. The coefficients $c_0 - c_6$ for the said platform are listed in Table ~\ref{tab:table_coeff_1} in terms of a suitable energy scale $\epsilon$ \cite{Poddubny2019}.  

\begin{table}[h]
\caption{\label{tab:table_coeff_1}%
Values of the coefficients $\{c_0-c_6\}$ in Eq.\ref{eq: Ham_haldane} for Rydberg excitons in cuprous oxide.}
\begin{tabular}{ l @{\qquad} c }
\toprule
\textrm{Coefficients}&
\textrm{Value/$\epsilon$}\\
\colrule
$c_0$ & -5.58 \\
$c_1$ & 9.53 \\
$c_2$ & -8.97 \\
$c_3$ & 1.27 \\
$c_4$ & 6.59 \\
$c_5$ & -3.18\\
$c_6$ & 5.04\\
\botrule
\end{tabular}
\end{table}

\begin{figure}[ht!]
\centering 
\includegraphics[width=3.55 in]{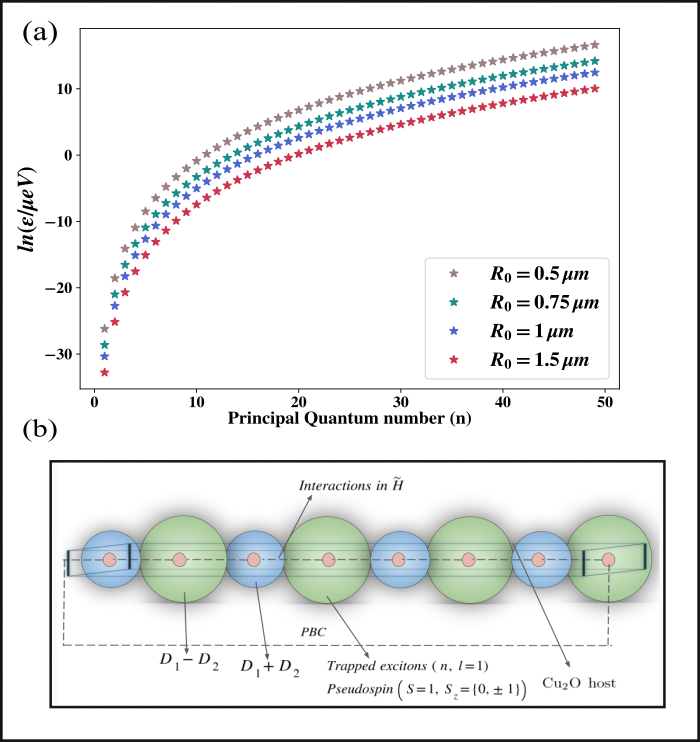}
\caption{(a) $log(\epsilon)$ ($\epsilon$ is expressed in $\mu eV$) vs principal quantum number of the Rydberg exciton state $n$ at various inter-exciton distance $R_0$ (in $\mu m$) (b) A schematic for $N = 8$ spins of trapped excitons in Cu$_2$O, in a state characterized by $(n,l=1)$ wherein $n$ is the principal and $l$ is the azimuthal quantum number of the Rydberg excitonic quantum state. The scheme assumes local modulation to create site-dependent anisotropies with $D_1 \ge D_2$ and $(D_1, D_2) > (0,0)$ which ensures lower exciton radius (blue) for tightly trapped sites (labeled as $D_1 + D_2$)}
\label{fig: fig_scheme_setup}
\end{figure}

Fig.~\ref{fig: fig_scheme_setup}(a) shows that the energy scale $\epsilon$ escalates steeply with increasing principal quantum number $n$ ($\propto n^{11}$) for the Rydberg excitons at a fixed $R_0$, i.e. the inter-exciton distance. This quantity decays with increasing $R_0$ ($\propto R_0^{-6}$ ) due to the reduction in the van der Waals interaction between the excitons. 

\begin{figure*}[ht!]
  \centering 
\includegraphics[width=6 in]{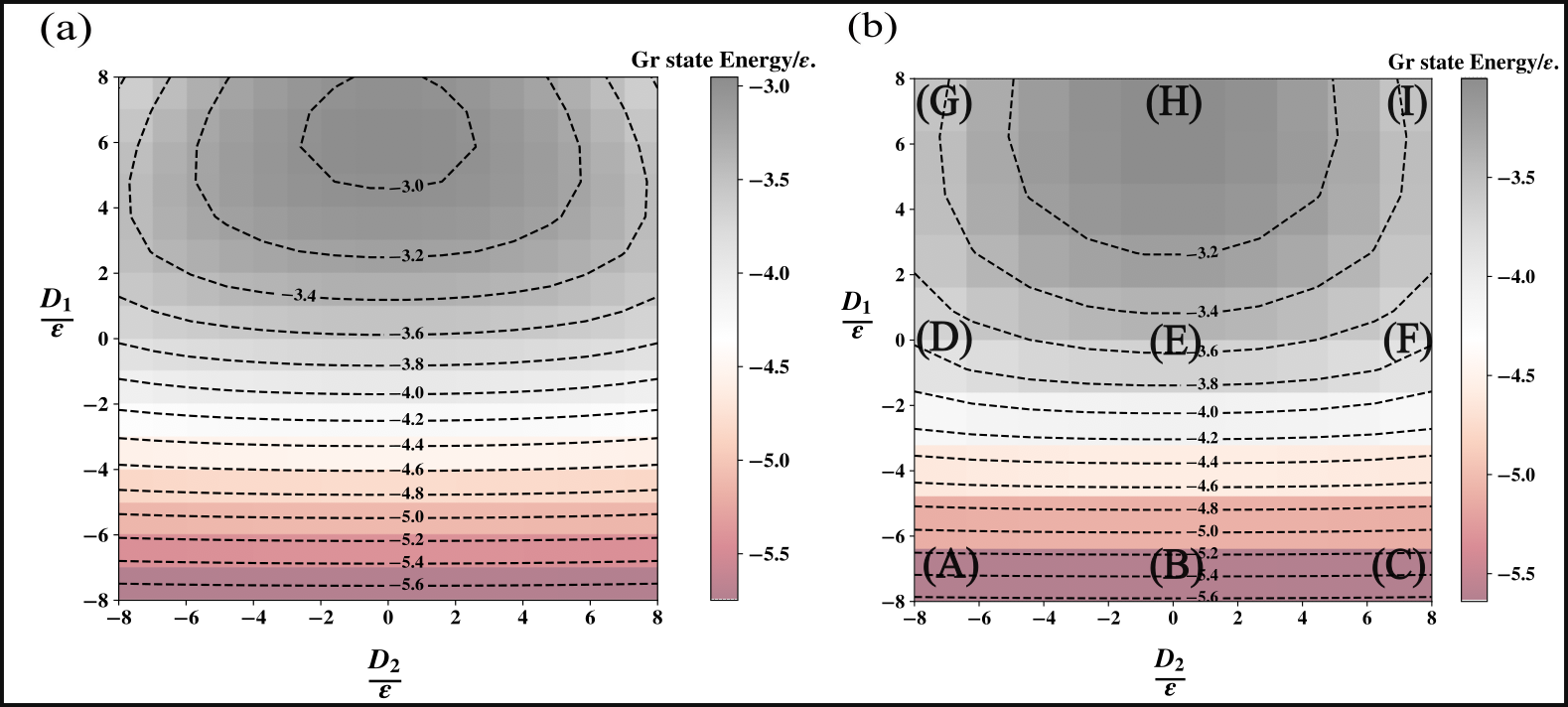}
\caption{Ground state energy of the Hamiltonian defined in Eq.~\ref{eq: Ham_anisotropy} as a function of local anisotropy profile $(D_1, D_2)$ in units of the energy scale $\epsilon$ for (a) $N=4$ (b) $N=16$ obtained from DMRG calculations. The regime near $|D_1|, |D_2| \approx (0,0)$ is the Haldane phase whereas for $D_1, D_2 \ll 0$ we have the anti-ferromagnet and for $D_1,D_2 \gg 0$ it is the large-D-limit disordered phase. The approximate position of the points $(A, B, C, D, E, F, G, H, I)$ defined in the text are demarcated in panel (b). The points as $(D_1, D_2)$ are $[A\equiv(-8,-8), B \equiv (-8,0), C \equiv (-8,8), D \equiv (0,-8), E\equiv (0, 0), F \equiv (0, 8), G \equiv (8,-8), H \equiv (8, 0), I \equiv (8,8)]$. }
\label{fig: E_fig_Gr_4_16}
\end{figure*}


Herein we further investigate the same spin model with locally-modulated the on-site energy as $\textbf{D} = \{D_j = D_1 + (-1)^j D_2\:|\: \forall \: j \in \mathbf{Z}_N, (D_1, D_2) \in \mathbf{R}^2\}$. The resultant spin system is similar to a Creutz ladder~\cite{Creutz1999, Alaeian2019, Huang2021} with two mutually interacting spin graphs with different local anisotropies characterized by $(D_1, D_2)$ (in units of the scale $\epsilon$) in an overall Hamiltonian of the form
\begin{eqnarray}
\tilde{H} = H + \sum_j D_j S_j^{z^2} \, . \label{eq: Ham_anisotropy}
\end{eqnarray}
The local anisotropy profile for a particular realization of $(D_1, D_2)$ is schematically represented in Fig.~\ref{fig: fig_scheme_setup}(b).This periodic modulation can be thought as two interlaced 1D exciton traps with different depth. Such 1D arrays can be realized using different approaches, e.g. via a periodic arrangement of static stressors~\cite{Kruger2018} or surface acoustic waves~\cite{Mingyun2022} to locally deform the crystal lattice and trap the excitons, or using spatial light modulators to create excitons at desired locations~\cite{Tao2022}, or through the optical confinement~\cite{Alloing2013}. 

Since $c_1 > 0$, the ground state of Eq.~\ref{eq: Ham_haldane} has an anti-ferromagnetic character ~\cite{Poddubny2019}, the predominance of which can be conveniently tuned using a local anisotropy profile as in Eq.~\ref{eq: Ham_anisotropy}. The modulated potential considered in this case serves to probe deeper into the physics of the model by demarcating the regime wherein such spin phases can be stabilized at will including the topologically non-trivial Haldane phase thereby highlighting its robustness. Further, it provides a pathway to transition from trivial phases to topological ones by controlling experimentally-accessible parameters by traversing a richer phase diagram hitherto unexplored.

Various phases would be characterized by one-body properties like local magnetization and von Neumann entropy followed by many-body properties like N\'eel correlation~\cite{Neel1948} and long-range string correlations~\cite{Affleck1987}. In the next section, we first explicate the classical resources used for the simulation and then present a linear-scaling quantum algorithm designed for the study of such systems.
%
%
\section{Classical Simulations}~\label{sub: DMRG}
The classical simulations of the spin model have been performed using the DMRG algorithm introduced by White~\cite{White1992} and later applied to many physical~\cite{PhysRevB.103.195122, PhysRevLett.97.110603,Stoudenmire2012, PhysRevB.94.045111} and chemically relevant problems~\cite{Chan2011, DMRG_exc_state, Baiardi2020}. The algorithm is extremely popular to circumvent the exponential scaling of many-body Hilbert spaces~\cite{Schollwock2011}, especially in 1D which is also the scope of this work. A detailed overview of such algorithms can be found elsewhere~\cite{Schollwock2011,Fishman2020}.


We used the Julia version of ITensor as well as the DMRGPY wrapper~\cite{DMRGPY} from ITensor library for computation with maximum bond dimensions of the required Matrix Product State (MPS) set to be 250 for each sweep. 100 sweeps are used over the linear array with a cutoff energy of convergence at $10^{-9}$ per sweep. For excited state computations, a weight parameter $\omega$ to penalize the overlap with the ground state is used  with steadily decreasing noise profile $\le 10^{-6}$ for a crisp convergence. 

Fig.~\ref{fig: E_fig_Gr_4_16}(a),(b) shows the ground state energy profile in the $(D_1, D_2)$ plane subject to PBC for two chain sizes $N=4$, $N=16$, respectively. We see that the ground state energy is lower for smaller values of $D_1$ irrespective of the modulation depth $D_2$ (in the range studied) and rises when $D_1$ is escalated.  We analyze the properties of the spin chain at the designated points (A-I) marked in Fig \ref{fig: E_fig_Gr_4_16}(b) for $N=16$. (Similar trends for different chain length can be observed.) Using spin correlations, in the next section we discuss the implications of this result and show that for $(|D_1|,|D_2|) \approx (0,0)$, e.g. point E, there is a distinct Haldane phase whereas for $(D_1, D_2) \ll 0$, e.g. point A, we have an anti-ferromagnet, and for $(D_1,D_2) \gg 0$, e.g. point I, we have the large $|D|$-limit disordered phase~\cite{PhysRevB.67.104401,Hida2005,PhysRevB.96.060404,PhysRevLett.98.047205}.
%
\begin{figure*}[ht!]
  \centering 
\includegraphics[width=6.8 in]{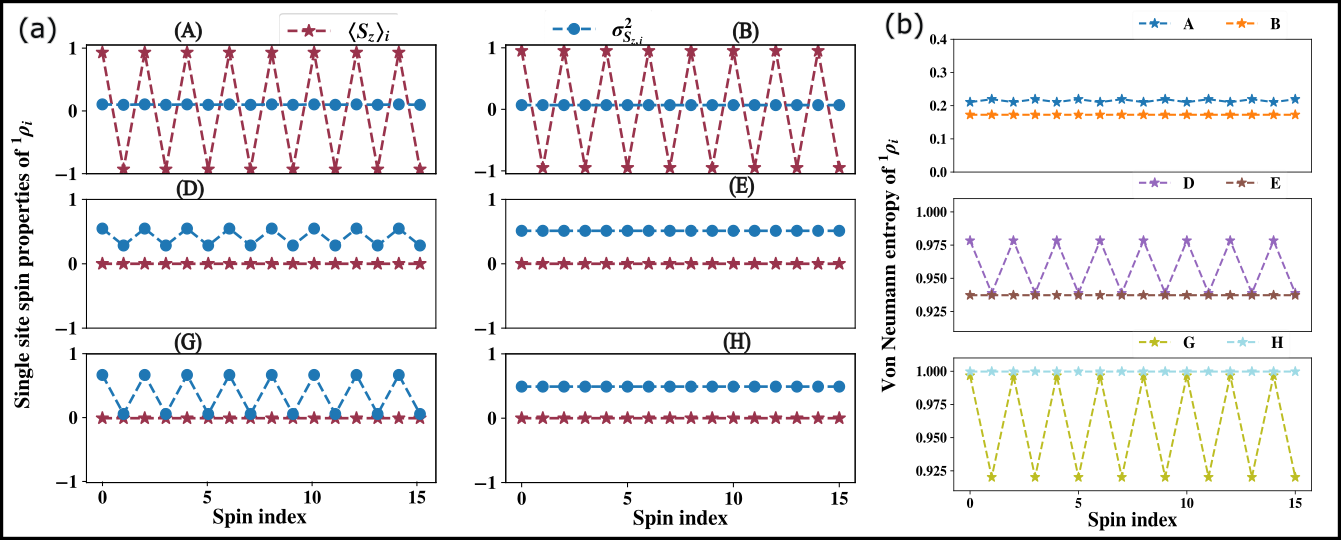}
\caption{(a) The statistical properties of $S_{z,i}$ along the chain (single-site magnetization about the z-axis) for the spin chain with $N=16$ in the ground state of the Hamiltonian defined in Eq.~\ref{eq: Ham_haldane} for different parameter regimes of $(D_1, D_2)$ corresponding to the points (A,B,D,E,G,H) in Fig.~\ref{fig: E_fig_Gr_4_16}(b). The two properties being displayed are single-site average magnetization $\langle S_z \rangle_i$ (red stars) and the associated variance $\sigma_{S_{zi}}^2$ (blue circles). The top panel (A,B) admits a non-zero and oscillatory $\langle S_z \rangle_i$ behavior with a nearly zero single-site variance indicating a dominant single configuration in the many-body state. The remaining panels corresponding to (D,E,G,H) have vanishing $\langle S_z \rangle_i$ with a non-zero variance indicating a many-body state comprising a superposition of many computational configurations.
(b) Single-site von Neumann entropy (expressed in units of $log(3))$ defined as $-Tr(^1\rho_i ln(^1\rho_i))$ for the ground state of the Hamiltonian in Eq.~\ref{eq: Ham_haldane} at $N=16$ for different modulation parameters $(D_1, D_2)$ corresponding to the points discussed on the left panels.}
\label{fig_Gr_one_body_16}
\end{figure*}
\begin{figure*}[ht!]
  \centering 
\includegraphics[width=5.0 in]{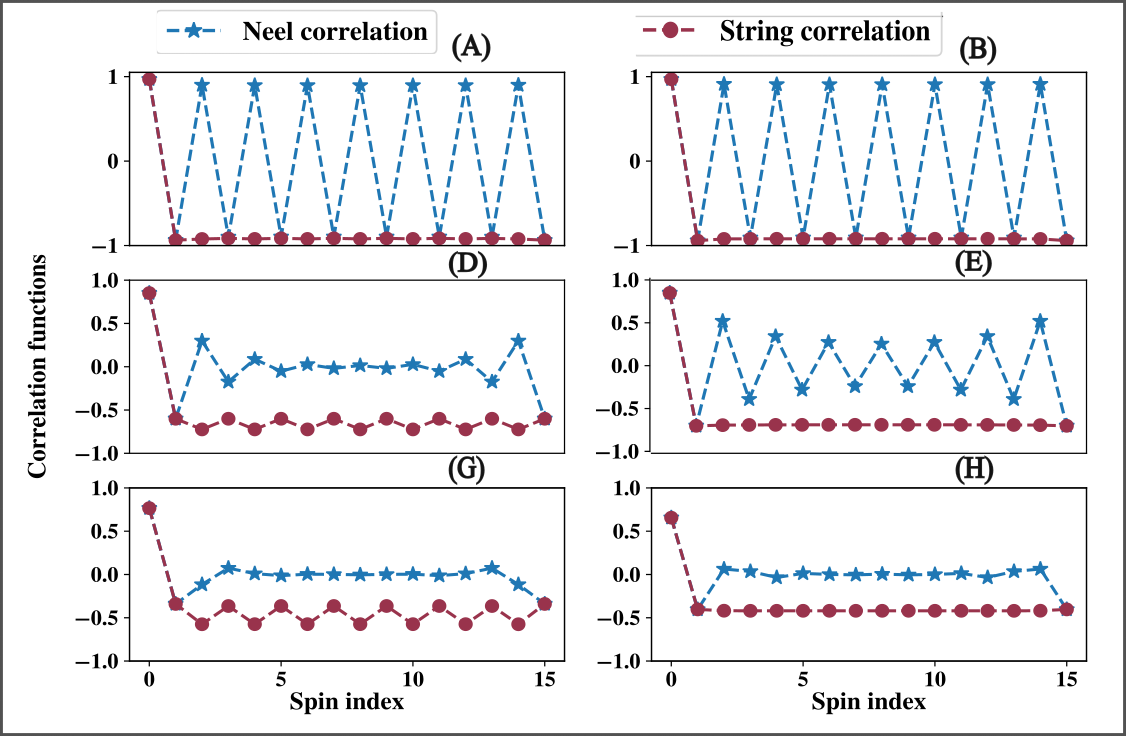}
\caption{
The ground-state pair-wise N\'eel order (blue stars) and the long-range string (red circles) correlation for $N = 16$ and different modulation parameters $(D_1, D_2)$, corresponding to the points (A,B,D,E,G,H) shown in Fig.~\ref{fig: E_fig_Gr_4_16}(b), as a function of site indices along the chain.}
\label{fig_Gr_two_body_16}
\end{figure*}
%

%


%
%
\subsection{Single and Many-body properties}~\label{sub: correlation}
To determine different phases obtainable in this periodically-modulated spin-1 chain we study certain order parameters like local magnetization ($\langle S_z^{(i)} \rangle$), its associated variance $\sigma_{S_{z,i}}$, and the von Neumann entropy of the reduced single-spin density matrix $^1\rho_i$ along the chain as well as the pair-wise N\'eel ($C_N(i)$) and the long-range string ($C_S(i)$) correlation defined as
\begin{eqnarray}~\label{eq:neel order}
    C_N(i) =  (-1)^i \braket{S_0^z S_i^z}\, ,
\end{eqnarray} 
and 
\begin{equation}~\label{eq:string order}
    C_S(i) = \braket{S_0^z e^{i\pi \sum_{p = 1}^{i-1} S_p^z} S_i^z}\, .
\end{equation}
The local average of of the magnetization $\langle S_z^i \rangle$ is plotted in Fig.~\ref{fig_Gr_one_body_16}(a) and the von Neumann entropy is shown in Fig.~\ref{fig_Gr_one_body_16}(b) respectively for points ($A,B,D,E,G,H$) defined in Fig.~\ref{fig: E_fig_Gr_4_16}(b). Since a similar local anisotropy profile is established in ($C, F, I$) as in ($A, D, G$) respectively with the odd and the even sites swapped (see Eq.~\ref{eq: Ham_anisotropy}) the former points are omitted from further discussion.

We see that for points $(D,E, G,H)$, the magnetization $\langle S_z^i \rangle$ at each site, depicted with red stars, is predominantly zero with a non-vanishing variance of $S^z_i$, locally. This essentially is a consequence of attaining a many-body quantum state as a superposition of many contributing configurations. However, that is not the case for $(A,B)$ due to a preference of one of the symmetry-broken configurational states as is indicated by the low on-site variance of $S^z$.

This is further bolstered in the corresponding von Neumann entropy profile presented in Fig.~\ref{fig: E_fig_Gr_4_16}(b) calculated from the reduced density matrix at each site ($^1\rho_i$). As can be seen the entropy attains a low value for $(A,B)$ due to the preference of one of the uncorrelated configurations leading to a pure one-site reduced density matrix ($^1\rho_i$) at all sites $i$. The single-site entropy for other points (D-E, G-H) however, are near maximal indicating an almost maximally-mixed reduced density matrix ($^1\rho_i$) at each site arising from a many-body state as the superposition of computational basis states.


To determine the phase of the spin chain at different points we investigate the correlations as shown in Fig.~\ref{fig_Gr_two_body_16} where the blue stars show the N\'eel order and the red circles indicate the string correlation. For points ($A,B$) wherein $D_1 = -8$, the N\'eel correlation is sustained and non-decaying irrespective of $D_2$. The modulating Hamiltonian, i.e. the second term in Eq.~\ref{eq: Ham_anisotropy}, leads to a stabilized ground state where sites with $S_p^z=\pm 1$ are favored over $S_p^z = 0$. 




In the disordered phase at $D_1 = 8$ however, e.g. $(G,H)$, the local anisotropy at each site is positive and hence unlike the previous case, configurations with $S_p^z=0$ are favored due to the non-negative penalty of $S_p^z=\pm 1$ sites enforced by the local modulation. As a result, many different configurations contribute significantly to the ground state and the overall anti-ferromagnetic order is lost with the lack of exclusive dominance from N\'eel-ordered configurations. 

When $D_1 = 0$, e.g. at $(D,E)$, the local anisotropy is $\pm 8$ for alternating spins and hence partial remnants of anti-ferromagnetic order as for $D_1 = -8$ damped quickly due the sites with local $S_p^z=0$ being favored owing to the non-negative penalty as in ($G,H$). We also see a non-decaying string correlation in points $(D,E)$ of Fig.~\ref{fig_Gr_two_body_16}. It must be emphasized that due to the nature of the exponential term in Eq.~\ref{eq:string order}, which can contribute $-1$ for sites with $S_z^p= \pm 1$ and $1$ for sites with $S_z^p = 0$, the string correlation remains negative for all points $(A-H)$, except at $j=0$. The said order between site indexed at `0' and at `j' can only acquire non-zeros values for configurations wherein the terminal spin $(S_z^0,S_z^j)$ are in any one of the four states $(\pm 1, \pm 1)$ since $S_z =0$ on either spin would destroy the correlation order defined in Eq.~\ref{eq:string order}. The non-negativity of string order can be partially explained using the number of contributing configurations atleast for the case of spin indices at the terminus of the excitonic array i.e. $j=N-1$. For all such configurations in which the terminal spins are of opposite spin-type, the intermediary spins that contribute to the exponential term in the string order must have $\sum_{p=1}^{j-1} S_z^p = 0$ to ensure an overall $\sum_j S_z^j = 0$ which is the symmetry of the ground as well as the first excited state (cf. Appendix~\ref{app: symmetry}). Therefore, the total number of such configuration is proportional to $2(\sum_{i=1}^{(j-1)/2} {j-1 \choose i}{j-2 \choose i} + 1)$, all having the string correlation of $-1$. On the other hand, when the terminal spins are in the same spin state for which the string order is $+1$, the intermediary spins in such configurations must have $\sum_{p=1}^{j-1} S_z^p = \pm 2$ which is proportional to a total count of $2({j-1 \choose 2}\sum_{i=1}^{(j-3)/2} {j-3 \choose i}{j-4 \choose i} + 1)$. Since the latter number is lower than the former one, the string order has an overwhelming majority of configurations wherein it attains a negative value. Furthermore, for ($A,B$), the dominant configuration(s) are of the first kind with terminal spins at $(\pm 1, \mp 1)$ which explains their larger string correlations compared to other points (cf. Appendix~\ref{app: config}).

The above discussion, together with the decaying N\'eel order and the sustenance of string correlation at points $(D, E, F)$ (especially for $E$) supports the emergence of the topological Haldane phase for this trap configurations. At point $E$ there is no perturbation hence we recover the properties of the Haldane phase discussed in~\cite{Poddubny2019}. In the next section we discuss how such spin-1 system can be mapped to an RBM based off spin-1/2 models with a hybrid-algorithm linearly scaling in qubit numbers with the chain length $N$ thereby allowing it to be studied on NISQ era quantum computers.



%
\section{Quantum Simulation Using Restricted Boltzmann Machine (RBM)}~\label{sub: RBM_sim}
Quantum machine learning algorithms based on parameterized quantum circuits have begun to gain enormous attention for understanding the physical properties of atomic and molecular systems~\cite{mehta2019high, sajjan2022quantum,10.1007/978-3-030-29407-6_5}. One such network, commonly known as the Restricted Boltzmann Machine~\cite{Melko2019a} has been employed successfully as a neural-network ansatz for a quantum state\cite{https://doi.org/10.48550/arxiv.2208.13384} in many different applications like in fermionic assemblies~\cite{Strong_corr_RBM,choo2020fermionic, ChNg2017}, anyonic symmetries~\cite{RBM_anyons_symm}, supervised phase-classification tasks~\cite{Ciliberto2017, Carrasquilla_2017}, in dynamical evolution~\cite{PhysRevResearch.3.023095} and in chemistry~\cite{sajjan2021quantum, sureshbabu2021implementation, Xia_2018, RBM_molecule}. The network is a universal approximator for any probability density~\cite{Roux_RBM,Melko2019a} and can efficiently simulate a volume-law entangled state even when sparsely parameterized~\cite{PhysRevX.7.021021}. It has been provably demonstrated that evaluating the full distribution classically would require exponential resources as long as polynomial hierarchy is not collapsed~\cite{Long2010}, however recently some of the authors have reported a quantum algorithm training the said ansatz to efficiently simulate the ground and excited state properties of materials and molecules with quadratic resources~\cite{sajjan2021quantum}.
\subsubsection{The Model}
\begin{figure}[ht!]
  \centering 
\includegraphics[width=3.5 in]{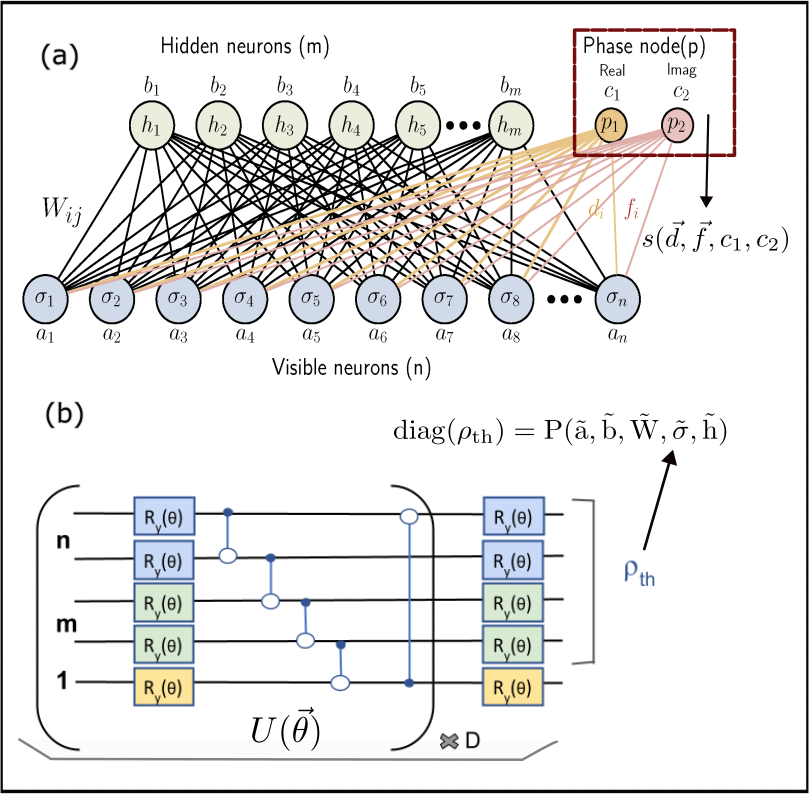}
\caption{(a) The scheme for the network $G$ used in this manuscript is made from the combination of the usual RBM network \cite{Fischer2014, Hinton504, ACKLEY1985147, 10.1162/089976602760128018, Melko2019a}  and two extra neurons (demarcated in red as Phase node) to account for the phase of the wave function. (b) The quantum circuit that has been used for variational training for generating the amplitude field of the RBM part of the network $G$ is shown for a specific case of $n=m=2$. The circuit is labelled as a parameterized unitary $U(\vec{\theta})$. The neurons of visible node and hidden node registers in (a) are replaced by qubits which are acted on by single-qubit unitaries with the respective color-coding maintained as in (a). The circuit also requires an ancillary qubit (single-qubit unitary denoted by yellow). In general the circuit requires $O(m+n)$ qubits and $O(D(m+n))$ gates where $D$ is the depth of the ansatz }
\label{fig_scheme_QRBM}
\end{figure}

The specific description of the network can be found in Fig.~\ref{fig_scheme_QRBM}(a).
The core idea involves encoding the target state of the desired Hamiltonian (to be called the driver system) into a neural-network architecture that resembles a bipartite spin graph $G$ with tunable connectivity. Formally, the graph can be envisioned as $G=(V,E)$. The set of vertices $V$ (henceforth to be called neurons) is further divided as $V = \{\nu\}_{i=1}^{m+n} = \{\sigma\}_{i=1}^{n} \bigcup \{h\}_{j=1}^m \bigcup \{p\}_{k=1}^2$ with $(m,n) \in \mathcal{Z_{++}}$. The neurons of the type $\Vec{\sigma} \in \{1, -1 \}^n$ (shown in blue in Fig.~\ref{fig_scheme_QRBM}(a)) collectively form the visible node register. For a given Hamiltonian matrix $H \in \mathcal{C}^{d \times d}$ of the driver system, the number $n$ is chosen such that $n = \lceil log_2 (d) \rceil$. The neurons of the type $\Vec{h} \in \{1, -1\}^m$ (shown in green in Fig.~\ref{fig_scheme_QRBM}(a)) constitute the hidden node register. This number $m$ can be chosen arbitrarily to accomplish the desired accuracy threshold (Here, $m \sim n$). The role of the hidden node register is to escalate the expressibility of the generative model by increasing the number of controllable parameters and inducing higher-order hidden correlation among the spins of the visible register~\cite{Hinton504, Fischer2014, Melko2019a}. Two neurons of the type $p_k \in [-1,1]$ (shown in orange and red in Fig.~\ref{fig_scheme_QRBM}(a) constitute the phase nodes whose role will be discussed, shortly~\cite{kanno2019manybody}. 

It is also assumed that all possible pairs of neurons $\sigma_i$ and $h_j$ are interconnected. Further, each $\sigma_i$ shares edges with the two neurons $\{p\}_{k=1}^2$, too. Collectively, all such edges are labeled as $\{e_{ij}\}_{i=1,j=1}^{n,m+2}$ and constitute the set $E$ where $|E| = mn + 2n$. 
Associated with the vertices $\nu_i \in V$ we define a $bias$ vector $(\vec{a}, \vec{b}, \vec{c}) \in \mathcal{R}^{m+n+2}$ where $\vec{a}=\{a\}_{i=1}^{n}$ is for neurons in the visible-node register $\vec{\sigma}$,  $\vec{b}=\{b\}_{j=1}^{m}$ for neurons in the hidden-node register $\vec{h}$, and $(c_1,c_2)$ for the two neurons $\{p\}_{k=1}^2$ in the phase-node register. Similarly, associated with $e_{ij} \in E$ where $i \in \lceil n \rceil$ and $j \in \lceil m \rceil$ , we define a $weight$ matrix $\vec{W} \in \mathcal{R}^{n \times m}$. Together with the tunable parameters $(\vec{a}, \vec{b}, \vec{W})$ we define the energy function $\mathcal{E}(\vec{a}, \vec{b}, \vec{W}, \vec{\sigma}, \vec{h})$ as
\begin{align}
    \mathcal{E}(\vec{a}, \vec{b}, \vec{W}, \vec{\sigma}, \vec{h}) = a^T\sigma + b^T h + h^T W \sigma\, .
    \label{eq: Ising_energy}
\end{align}
This subset of the network encodes a probability distribution which is essentially the classical thermal state of the corresponding Ising spin Hamiltonian in Eq.~\ref{eq: Ising_energy} defined as follows~\cite{10.1088/978-0-7503-3843-1ch2,RevModPhys.39.883,PhysRevB.94.165134,torlai2018neural}
\begin{align}
P(\Vec{a}, \Vec{b}, \Vec{W}, \vec{\sigma},\vec{h}) &= \frac{e^{-\mathcal{E}(\vec{a}, \vec{b}, \vec{W}, \vec{\sigma}, \vec{h})}}
{\sum_{\{\sigma h\}}e^{-\mathcal{E}(\vec{a}, \vec{b}, \vec{W}, \vec{\sigma}, \vec{h})}}\, . \label{eq:rbm_dist}
\end{align}
This subset of $G$ thus constitutes the usual RBM network. 

Any realization of spin configuration $(\vec{\sigma}, \vec{h})$ of the combined registers of $(m + n)$ spins are sampled from the distribution in Eq.~\ref{eq:rbm_dist}. Since the parameter vector $(\vec{a}, \vec{b}, \vec{W})$ is real-valued, the functional form in Eq.~\ref{eq:rbm_dist}
mimics a probability distribution, only. To account for the phase of the target wavefunction, herein we encode the following complex-valued function within the neurons $\{p_k\}_{k=1}^2$ 
through the following activation function\cite{kanno2019manybody}
\begin{align}
s(\Vec{d}, \Vec{f}, \vec{c}, \vec{\sigma}) &= \tanh\left[(c_1 + \sum_{i}d_i\sigma_i) + i(c_2 + \sum_{i}f_i\sigma_i)\right]\, ,  \label{eq:phase_encod}
\end{align}
where $\Vec{d} , \Vec{f} \in \mathcal{R}^n$ are the $weights$ for the connections between $\{p\}_{k=1}^2$ and and $\{\sigma\}_{i=1}^{n}$ (see Fig.~\ref{fig_scheme_QRBM}(a)). The full ansatz thus consists of neurons of the RBM as well as the two neurons of the phase node with a complete set of parameter vector $X = (\vec{a}, \vec{b}, \vec{W}, \vec{d}, \vec{f}, c_1, c_2)$ that can be adjusted variationally during the training process.
\subsubsection{The Outline of the Algorithm and Results}
The purpose of using the aforesaid generative network is to construct an ansatz for the amplitude and the phase of the target state wavefunction for the Hamiltonian defined in Eq.~\ref{eq: Ham_anisotropy} as follows
\begin{align}
    \psi(\Vec{X}) = \sum_{\sigma}  \sqrt{\sum_{\vec{h}}P(\Vec{a}, \Vec{b}, \Vec{W}, \vec{\sigma}, \vec{h})}s(\Vec{d}, \Vec{f}, \vec{c}, \vec{\sigma})|\sigma_1\sigma_2...\sigma_n\rangle  \, , \label{wavefn}
\end{align}
wherein $|\sigma_1, \sigma_2, ....\sigma_n \rangle$ represents the spin configurations of the visible node register $\{\sigma\}_{i=1}^{n}$ which forms the  computational basis in which the target wavefunction is resolved. 
The parameter vector $\vec{X} = (\vec{a}, \vec{b}, \vec{c}, \vec{W}, \vec{d}, \vec{f})$ is tuned until the energy of the ansatz state is variationally minimized. The wavefunction in Eq.~\ref{wavefn} obtained at the end of this training process has a large overlap with the target state. 

Recently, the authors have developed an algorithm to retrieve the distribution directly using a quantum circuit~\cite{sajjan2021quantum} with a circuit width, a number of gates, and parameter count of $O(m\times n)$ for a visible (hidden) node register of $n(m)$ qubits. Since for S = 1, conversion of the problem to qubit degrees of freedom already leads to wasteful circuit widths, using the aforesaid quadratic scaling algorithm for this system would be inefficient. To this end here we develop a \emph{linear} scaling algorithm $O(m + n)$ which also retrieves the required distribution function based on an arbitrary Gibbs-state preparation scheme as described below. The overarching theme of our algorithm involves a double optimization protocol. The outer-optimization involves tuning the parameter vector $\vec{X}$ whereas the inner optimization involves constructing the Gibbs state given an $(\vec{a}, \vec{b}, \vec{W})$ defining a connected Ising spin system. %
\begin{itemize}
    \item The algorithm starts by randomly assigning values to the parameter vector $\vec{X} = (\vec{a}, \vec{b}, \vec{W}, \vec{c}, \vec{d}, \vec{f})$ by sampling from a uniform distribution within [-0.02, -0.02] to avoid the problem of vanishing gradient~\cite{Xia_2018}. If training with random initialization is not successful within the desired accuracy threshold (say $\eta$), an initial parameter set of a previously converged point in a similar optimization problem is used as the initial guess. This process is called warm starting\cite{sajjan2021quantum,sureshbabu2021implementation} and is routinely used in machine learning for non-convex functions as is the case here.
    
    \item The next step is using the $(\vec{a}, \vec{b}, \vec{W})$ set to construct a quantum circuit that would return the thermal state $\rho_{th}(\vec{a}, \vec{b}, \vec{W})$ of $(m +n)$ qubits corresponding to the Ising energy function (defined in Eq.~\ref{eq: Ising_energy}). The probability distribution that encodes the amplitude field of the ansatz as defined in Eq.~\ref{eq:rbm_dist} is then retrieved from the diagonal elements of this thermal state. To construct the thermal state for the specific RBM energy function we follow the protocol designed in~\cite{PhysRevApplied.16.054035} using an ansatz $U(\vec{\theta})$ (see Fig.~\ref{fig_scheme_QRBM}(b)) with a circuit width of $O(m+n)$ and a depth of $D$. The depth $D$ has to be tuned by the user for attaining the accuracy threshold of interest. For this work, $D=3$ suffices for all examples. The protocol relies on minimizing the Helmholtz free-energy of $(m+n)$ qubits as the cost-function $F(\rho(\vec{\theta}, \vec{a}, \vec{b}, \vec{W}))$ with respect to $\vec{\theta}$ keeping $(\vec{a}, \vec{b}, \vec{W})$ fixed at the input value. The extra ancillary qubit acts as a digital bath to facilitate the formation of a mixed state for the register with $m+n$ qubits. (cf. Appendix~\ref{app:circuit_des} for more information about this step)
     
    \item With $P(\Vec{a}, \Vec{b}, \Vec{W}, \Vec{\sigma},\Vec{h})$ from the previous step of the algorithm, the phase function defined in Eq.~\ref{eq:phase_encod} was computed using the remaining part of the parameter set $(\vec{c}, \vec{d}, \vec{f})$. Combining the two, the wavefunction $\psi(\vec{X})$ can now be constructed and the cost-function for the outer loop of the optimization be evaluated. Since the objective is learning the eigenstates of the Hamiltonian $\tilde{H}$ in Eq.~\ref{eq: Ham_anisotropy} we use the following cost function for tuning the parameter vector $\vec{X}$
     \begin{align}
         &C(|\psi (\vec{X})\rangle, \tilde{H}, \hat{O}, \lambda) \nonumber \\
         &=\langle \psi| \tilde{H} |\psi \rangle + \lambda \langle \psi| (\hat{O}-\omega)^2 |\psi \rangle\, , \label{cost_fn_outer}
     \end{align}
     where $\lambda \in \mathcal{R}^{++}$ is the penalty set as a hyper-parameter. $\hat{O}$ is a user-defined symmetry operator ($[\tilde{H}, \hat{O}] = \hat{0})$ of the system whose eigenstate is desired with a specific eigenvalue $\omega$.
     The same prescription allows us to not only target specific states based on symmetry but also explore excited states by sampling the orthogonal complement of the ground state projection operator $P_{gr}$, i.e. by choosing $\hat{O} = P_{gr}$ and $\omega = 0$. One must note for the ground state $\lambda = 0$ is substituted in Eq. \ref{cost_fn_outer} as the usual minimization of energy  suffices.
     
     \item Once the cost-function $C(|\psi (\vec{X})\rangle, \tilde{H}, \hat{O}, \lambda)$ is evaluated in the outer loop, one can check for convergence (if $C(|\psi (\vec{X})\rangle, \tilde{H}, \hat{O}, \lambda) \le \eta$ where $\eta$ is the convergence threshold) or if maximum number of iterations has been completed. In either case, the resultant state $|\psi(\vec{X^*})\rangle$ and energy $\langle \psi (\vec{X^*)}| \tilde{H} |\psi(\vec{X^*}) \rangle$ is printed wherein $\vec{X^*} = \argmin_{\vec{X}}C(|\psi (\vec{X})\rangle, \tilde{H}, \hat{O}, \lambda)$.
     
     However, if either condition is not satisfied then the parameter vector $\vec{X}$ is updated using a gradient-based algorithm (ADAM optimizer~\cite{Kingma2015} is used in this case) with learning rate $\alpha_2 \in \mathcal{R}^{++}$ as 
     \begin{align*}
         \vec{X} \rightarrow \vec{X} - \alpha_2 \partial_{\vec{X}} C(|\psi (\vec{X})\rangle, \tilde{H}, \hat{O}, \lambda)\, .
     \end{align*}
     With the updated parameter set, one returns to the second step of the algorithm for the next iteration. The process is continued until the desired threshold convergence is reached or the maximum number of iterations set is exhausted.
    
\end{itemize}

\begin{figure}[ht!]
  \centering 
\includegraphics[width=3.5 in]{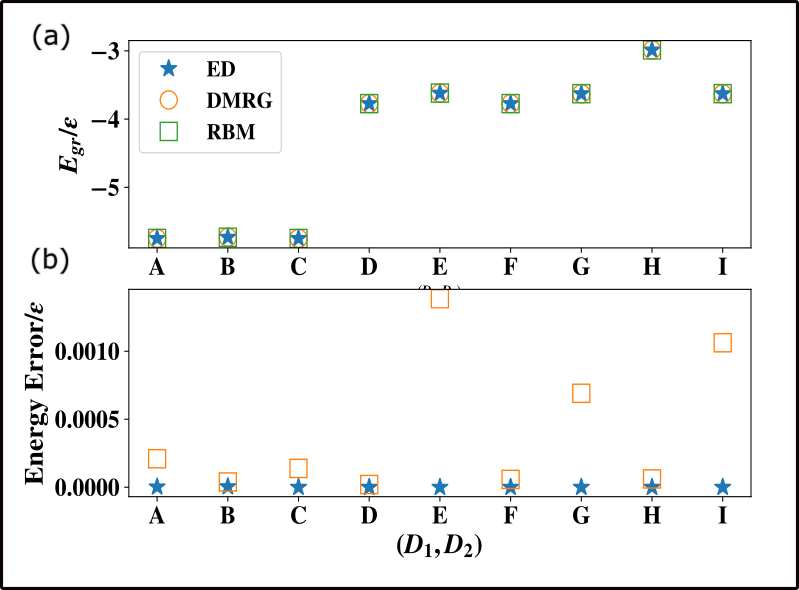}
\caption{(a) The ground state energy ($E_{gr}$) for $N=4$ spins in Eq.~\ref{eq: Ham_anisotropy} for points $(D_1,D_2)$ marked in Fig.~\ref{fig: E_fig_Gr_4_16}(b) computed by training the network $G=(V,E)$ abbreviated as RBM as well as DMRG and the exact diagonalization (ED). All results are expressed in units of $\epsilon$ as before. (b) The corresponding energy errors from Fig.~\ref{fig_Eerror_Gr_state}(a) for computations from both network $G$ and DMRG with respect to ED.}
\label{fig_Eerror_Gr_state}
\end{figure}

%
As a proof of concept, we now use the algorithm described to compute the ground states for a linear array of $N=4$ excitons in Eq.~\ref{eq: Ham_anisotropy} for different values of $(D_1,D_2)$ as indicated in Fig.~\ref{fig: E_fig_Gr_4_16}(b) under the PBC. The corresponding ground state energies and their associated errors are displayed in Fig.~\ref{fig_Eerror_Gr_state}(a) and (b), respectively. The energy errors are defined with respect to results of the exact diagonalization (ED) with a convergence threshold of $2\times 10^{-3}\epsilon$. Note that the choice of this threshold is motivated by the fact that it corresponds to a relative error percentage of $< 0.1 \%$ as can be established using the absolute values of the energy in Fig.~\ref{fig: E_fig_Gr_4_16}(a) and the obtained errors from Fig.~\ref{fig_Eerror_Gr_state}(b). The circuit described has been simulated using {\it statevector simulator} backend in Aer provider in Qiskit which corresponds to IBM's Quantum Information Software Kit (Qiskit)~\cite{aleksandrowicz2019qiskit}. 
For $N=4$, the total number of configurations accessible to the state-space would be $d=3^4=81$ which would require $n=7$, $m\sim 7$, and an additional ancillary qubit for the simulation of the circuit as described in the algorithm above. However, as the ground state is a singlet, symmetry restrictions allow us to disregard non-conforming configurations tapering the number of qubits to $(n=m=5)$. The total circuit width is $11$ which is already significantly higher than the usual simulation requirements with the same number of spins in S = $\frac{1}{2}$ case. For all the $(D_1,D_2)$ points studied, we have seen remarkable agreement with the exact results and DMRG calculations.

\section{Conclusion}~\label{sec: conclusion}
In this work we studied different topological and magnetic phases of a spatially-modulated 1D chain of spin-1 particles with the most general nearest-neighbor interactions realizable with optically-active Rydberg $p$-excitons in Cu$_2$O. Different phases have been categorized based on local spin magnetization and the von Neumann entropy as well as their pair-wise N\'eel and long-range string correlations. By studying the phases as a function of the local modulation we examined both the phase transitions as well as their robustness to local perturbations. 
Later, we proposed a quantum algorithm that allowed us to simulate the dynamics of such a system on a quantum computer using a hybrid  approach based on RBM.

As mentioned, strongly-interacting Rydberg excitons in Cu$_2$O are promising candidates for realizing analog quantum simulators in a solid-state platform. The ability to control the interaction strength is critical for investigating many-body systems and in this case the interaction mediated via Rydberg excitons can be controlled with the inter-particle spacing $R_0$ as well as the principal quantum number of the state, $n$. Therefore, Rydberg excitons provide a highly controllable and tunable tabletop solid-state platform with direct optical access to the individual sites for the coherent control as well as the readout. 
Since excitons are inherently out-of-equilibrium particles, an important follow-up study is to investigate the effect of drive and dissipation on the results presented in this work. Further, it would be interesting to explore the implications of such phases on the photoluminescence spectra of the exciton chain to examine whether the attainable magnetic and topological phases can be directly studies by laser spectroscopy techniques. 
Another interesting direction is the study of long-range interactions readily achievable with Rydberg excitons in such spin systems. It would be interesting to explore the generalized models with interactions beyond the nearest-neighbor and study their effects on the magnetic and topological phases obtained in this work. 
\section*{Acknowledgments}
The authors acknowledge the fruitful discussions with Stefan Scheel and Valentin Walther. S.K. and M.S. would like to acknowledge  the financial support of the U.S. Department of Energy, Office of Science, National Quantum Information Science Research Centers, and Quantum Science Center. H.A. acknowledges the Purdue University Startup fund. We acknowledge the use of IBM Quantum services for this work. The views expressed are those of the authors and do not reflect the official policy or position of IBM or the IBM Quantum team.

\bibliography{Main}


\end{document}


\preprint{APS/123-QED}

\title{Supplementary Information for Magnetic Phases of Spatially-Modulated Spin-1 Chains in Rydberg Excitons: Classical and Quantum Simulations}

\author{M. Sajjan}
\email{msajjan@purdue.edu}
\affiliation{Department of Chemistry, Purdue University, West Lafayette, IN 47907}

\author{S. Kais}
\email{kais@purdue.edu}
\affiliation{Department of Chemistry, Purdue University, West Lafayette, IN 47907}

\affiliation{Department of Physics and Astronomy, Purdue University, West Lafayette, IN 47907}

\author{H. Alaeian}
\email{halaeian@purdue.edu}
\affiliation{Elmore Family School of Electrical and Computer Engineering, Purdue University, West Lafayette, IN 47907}

\affiliation{Department of Physics and Astronomy, Purdue University, West Lafayette, IN 47907}

\maketitle

\widetext

\section{Symmetry}~\label{app: symmetry}
The Hamiltonian in Eq.1 in main manuscript can be written as 
\begin{equation}
    \hat{H} = \sum_{i = 1}^N H_{bond} (S_i,S_{i+1})\, .
\end{equation}

We define the total spin operator $\hat{S}_z$ as
\begin{equation}
    \hat{S}_z = \sum_{i = 1}^N \hat{s}_z^{(i)} \, ,
\end{equation}
where $\hat{s}_z^{(i)}$ is the i$^{th}$ site spin operator.

It is straightforward to see that $[\hat{H},\hat{S}_z] = 0$ implying that total spin is a conserved quantity of this Hamiltonian hence a good quantum number.

Another symmetry of the Hamiltonian is the permutation operator between sites (i,j) defined as follows

\begin{equation}
    \hat{p}_{ij} = (\vec{\hat{S}}_i \cdot \vec{\hat{S}}_j)^2 + (\vec{\hat{S}}_i \cdot \vec{\hat{S}}_j) - \hat{I}\, .
\end{equation}
For a chain of length $N$ if $j = N-i+1$, this transposition is another symmetry of the Hamiltonian, physically implying that the system energy remains unchanged upon the spin exchange in sites symmetric with respect to the chain center.

Using such permutation operators a parity operator can be defined as
\begin{equation}
    \hat{P} = \Pi_{i = 0}^{N/2} ~ \hat{p}_{i,N-i+1} \, .
\end{equation}

Clearly $\hat{P}$ is a $Z_2$ symmetry of $\hat{H}$ as $[\hat{H} , \hat{P}] = 0$ and $\hat{P}^2 = 1$.

$\hat{P}$ being a parity operator means that $\hat{P}\ket{\psi} = \pm \ket{\psi}$ for $\ket{\psi}$ being an eigenstate.
For the un-perturbed Hamiltonian the ground state and the excited states are mirror-symmetric and anti-symmetric corresponding to $\pm 1$, respectively.

For the spatially-modulated chain with the Hamiltonian in Eq.2, the parity is broken in general but the total spin operator $\hat{S}_z$ is still a symmetry of the total Hamiltonian meaning that even in this case the total spin is a good quantum number which is zero for the ground state, i.e. $\braket{\hat{S}_z} = 0$.

\section{Description of the quantum state in the configuration basis for various phases}~\label{app: config}

In this section we discuss the quantum state decomposition profile when resolved in the configuration basis (eigenbasis of local $S_{z_i}$) for the case of $N=4$ excitons in all the three phases studied. A similar profile exists for $N=16$ as well but the total number of configurations being $3^{16}$ makes it harder to interpret and visually characterize. This plot essentially explains the trends seen for $N=16$ in Fig.3(a) and (b) and Fig.4 of the main manuscript.

 \begin{figure}[ht!]
  \centering 
\includegraphics[width=6.8 in]{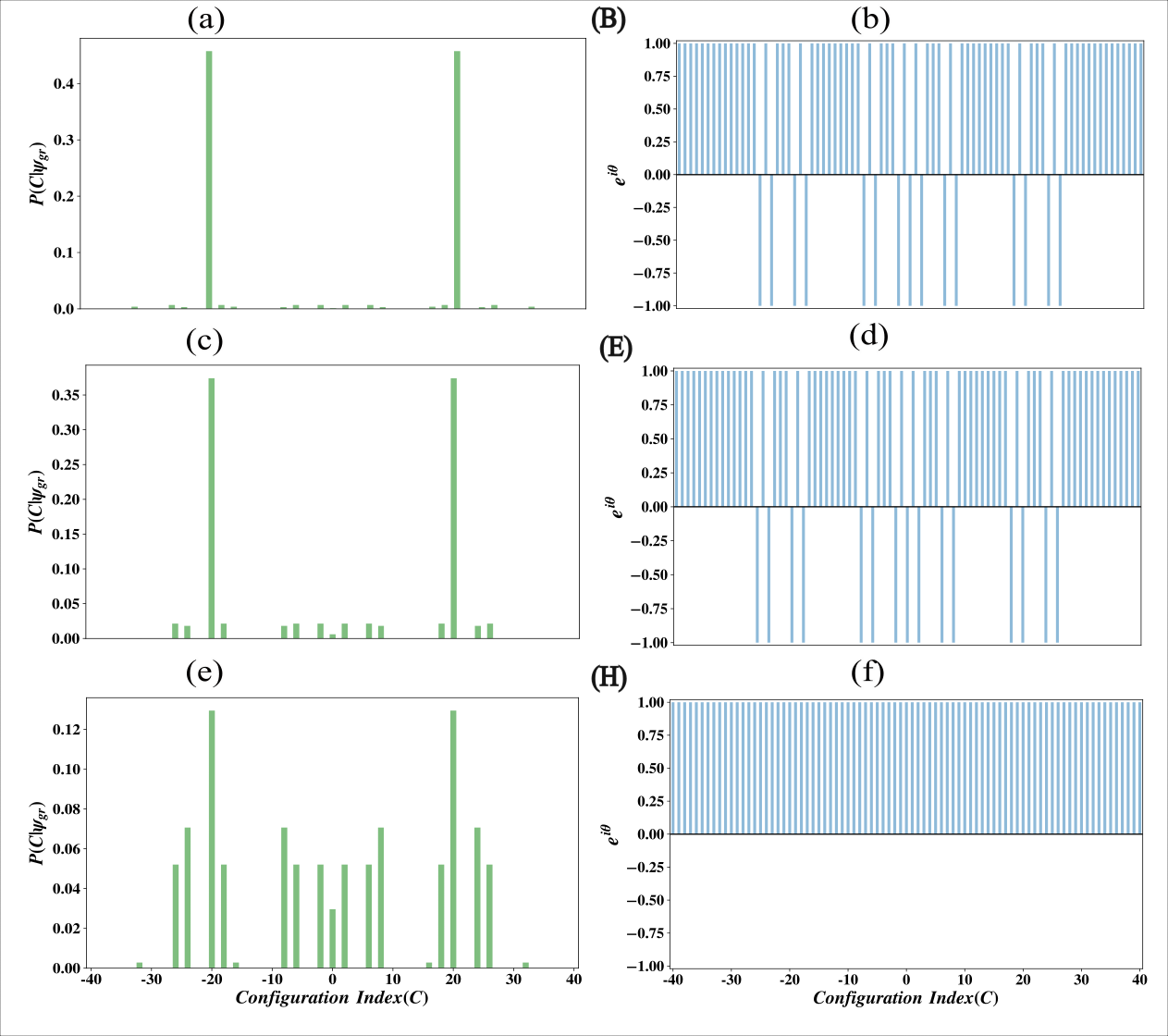}
\caption{(a) The probability distribution for the various configurations (configurations are strings containing $\langle S_z \rangle_i$ values of each excitonic site within the array) for the ground state at point $B=(-8,0)$ as marked in Fig.2(b) for $N=4$ excitons. The x-axis denotes the configuration index which is essentially a map from the space of balanced ternary to natural numbers. In other words for a configuration of $N=4$ of the kind $|i_3 i_2 i_1 i_0\rangle$, the corresponding index is computed as $\sum_{j=0}^{N-1} 3^j i_j$. The configuration indices for the two Neel ordered states ($|-1 1 -1 1\rangle$ and $|1 -1 1 -1\rangle$) in this scheme are $-20$ and $20$ respectively. Since $B$ is an anti-ferromagnet, we see near exclusive occupation of the two N\'eel-ordered bit-strings. Note that unlike in the case of $N=16$, for $N=4$, the ground state at $B$ is still a superposition state. However since the these states are degenerate, one can obtain any states in the corresponding sub-space including a symmetry-broken single state as seen for $N=16$. (b) The phase profile corresponding to (a) for various configurations at point $B=(-8,0)$. (c) The probability distribution for the various configurations for the ground state at point $E=(0,0)$ as marked in Fig.2(b) for $N=4$ chain. We see apart from the ordered states at indices $20(-20)$ as seen in (a), other configuration also contribute which leads to a decay of N\'eel order as seen for $N=16$ in Fig.4 or increased Von Neumann entropy or the variance of $S_{z_i}$ as seen in Fig.3(a),(b). (d) The  phase profile corresponding to (c) for the ground state at point $E=(0,0)$. (e) The probability distribution of various configurations for the ground state at point $H=(8,0)$ as marked in Fig.2(b). This corresponds to the disordered phase and hence is consequently populated almost equally by a vast array of different configurations. This leads to a complete lack of N\'eel order as seen for $N=16$ in Fig.4 or increased Von Neumann entropy and the variance of $S_{z_i}$ as seen in Fig.3(a-b). (f) The  phase profile corresponding to (e) for the ground state at point $H=(8,0)$.}
\label{fig_N_4_desc}
\end{figure}

\newpage
\section{Description of generation of thermal state using ansatz $U(\vec{\theta})$}
\label{app:circuit_des}

In this section we discuss the discuss the details of the circuit ansatz $U(\vec{\theta})$ and how the hybrid quantum algorithm discussed in Section IV of the main manuscript can be implemented on an actual quantum hardware.

Within the ansatz $U(\theta)$ shown in Fig.5(b), the $R_y(\theta)$ gates act to change the coefficients of the target state in the computational basis whereas the CNOT gate stores the parity of the respective qubits hence, is responsible for shuffling the coefficients among the computational basis states. The extra ancilla qubit (single qubit unitary of this ancillary qubit is marked in Fig.5(b) in yellow) acts as a digital bath. Coupling of the register of $(m+n)$ qubits with this ancilla through the cyclically arranged CNOT gates are responsible for the targeted decoherence of the initial state $|0\rangle \langle 0|_{m+n}$ into the thermal state $\tilde{\rho_{th}}(\vec{a}, \vec{b}, \vec{W})$. The depth of the circuit is $D$ and hence the number of single-qubit gates would be $(m+n+1)(D+1)$ and two-qubit CNOT gates would also be $(m +n +1)D$. 

The cost-function required to train the ansatz is the Helmholtz free energy $F(\rho(\vec{\theta}, \vec{a}, \vec{b}, \vec{W}))$ which is variationally mimimzed with respect to $\vec{\theta}$ keeping $(\vec{a}, \vec{b}, \vec{W})$ fixed. This is owing to the following equality between $F(\rho(\vec{\theta}, \vec{a}, \vec{b}, \vec{W}))$ and $S(\rho(\vec{\theta}, \vec{a}, \vec{b}, \vec{W}) || \rho_{th}(\vec{a}, \vec{b}, \vec{W}))$ which ensures that the state of minimum free energy is the requisite thermal state of the system~\cite{ruskai1990convexity}. 

    \begin{eqnarray}
        &F&(\rho(\vec{\theta}, \vec{a}, \vec{b}, \vec{W}))  =  E(\rho(\vec{\theta}, \vec{a}, \vec{b}, \vec{W}))
        - \beta^{-1} S(\rho(\vec{\theta}, \vec{a}, \vec{b}, \vec{W})) \nonumber\\
        &=& F(\rho_{th}(\vec{a}, \vec{b}, \vec{W}) + \beta^{-1}S(\rho(\vec{\theta}, \vec{a}, \vec{b}, \vec{W}) || \rho_{th}(\vec{a}, \vec{b}, \vec{W}))\, , \label{eq:free_energy_cost}
    \end{eqnarray}
wherein $E(\rho(\vec{\theta}, \vec{a}, \vec{b}, \vec{W})) = Tr (H_{is}(\vec{a}, \vec{b}, \vec{W}) \rho(\vec{\theta}))$ and $S(\rho(\vec{\theta}, \vec{a}, \vec{b}, \vec{W})) = -Tr(\rho(\vec{\theta}) ln \rho(\vec{\theta}))$ with $H_{is}$ being the Ising Hamiltonian operator obtained from quantizing the energy expression in Eq.5 in the main manuscript and $\beta$ is the usual inverse temperature chosen as a hyper parameter. The step involves computing gradient-based updates of the parameter set $\vec{\theta}$ with a learning rate $\alpha_1$ as
    \begin{align*}
          \vec{\theta} \rightarrow \vec{\theta} - \alpha_1 \partial_{\vec{\theta}} F(\rho(\vec{\theta}, \vec{a}, \vec{b}, \vec{W}))\, .
      \end{align*}
The step terminates either when the maximum set of iterations is reached or when $F(\rho(\vec{\theta}, \vec{a}, \vec{b}, \vec{W})) \le \epsilon$, the preset accuracy threshold.
Post-termination, the optimized vector $\vec{\theta^*}$ so obtained from the minimization encodes the desired state within the register comprising of $(m+n)$ qubits (see Fig.5(a-b)) as

\begin{eqnarray}
        \tilde{\rho_{th}}(\vec{a}, \vec{b}, \vec{W}) &=&
         U(\vec{\theta^*}) |0\rangle \langle 0|_{m+n} U^{\dagger} (\vec{\theta}^*) \nonumber \\
         \vec{\theta}^* &=& \argmin_{\vec{\theta}} F(\rho(\vec{\theta}, \vec{a}, \vec{b}, \vec{W}))\, ,
\end{eqnarray}
wherein $\tilde{\rho_{th}}(\vec{a}, \vec{b}, \vec{W}) \approx \rho_{th}(\vec{a}, \vec{b}, \vec{W})$. The required distribution in Eq.6  $P(\Vec{a}, \Vec{b}, \Vec{W}, \vec{\sigma},\vec{h}) = diag(\tilde{\rho_{th}}(\vec{a}, \vec{b}, \vec{W}))$.

\begin{figure}[ht!]
  \centering 
\includegraphics[width=5.0 in]{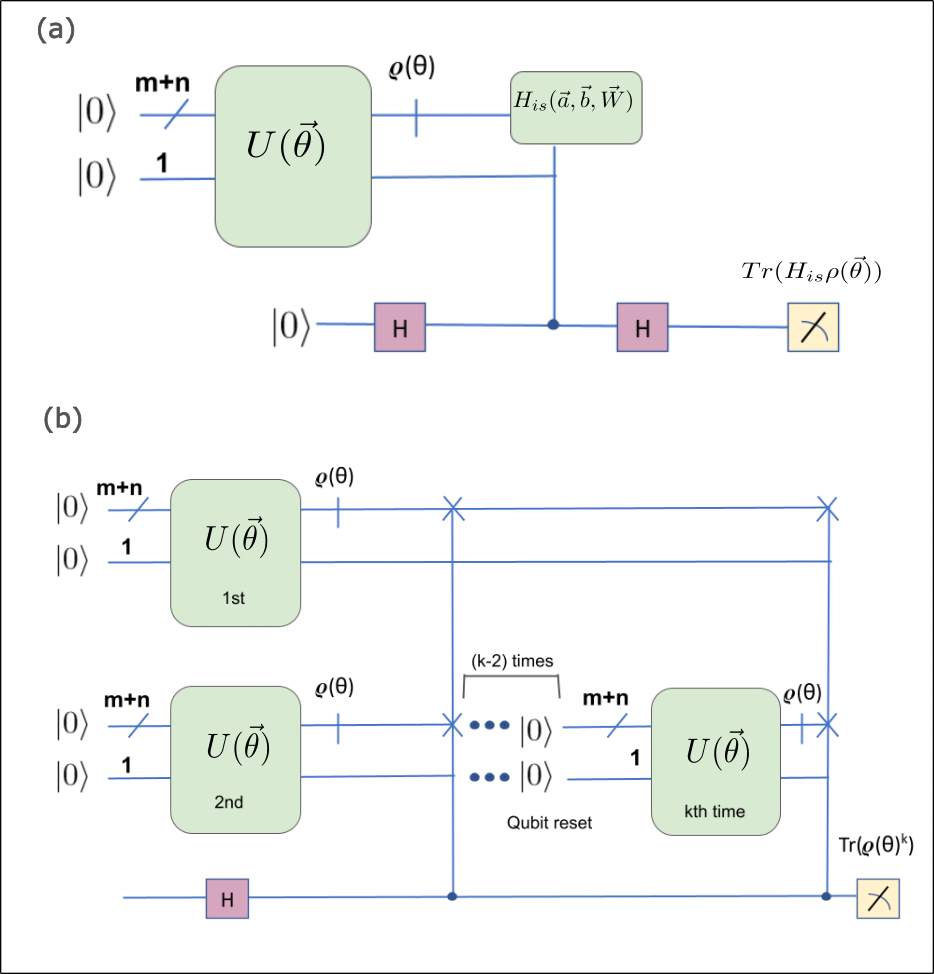}
\caption{(a) The Hadamard test for estimating the real part of average of any target unitary operator in the state prepared by the parameterized unitary $U(\vec{\theta})$. In general any target unitary can be used and the circuit as described is capable of returing only the real part of the average. Here for the target unitary we use the Pauli decomposed form of $H_{is}$ (obtained by quantizing Eq.5in main manuscript). Since each such Pauli string is both unitary and hermitian, the real part suffices to yield $Tr (H_{is}(\vec{a}, \vec{b}, \vec{W}) \rho(\vec{\theta}))$ where $\rho(\vec{\theta})$ is the state prepared by $U(\vec{\theta})$ (see Fig.5(b) in main manuscript) over (m+n) registers. In this scheme only the ancillary qubit is measured and the sampling probabilities are post-processed to get the required average. (b) The qubit-efficient version of the SWAP test as illustrated in Ref.~\cite{yirka2021qubit}. The circuit returns $Tr(\rho(\vec{\theta}, \vec{a}, \vec{b}, \vec{W})^k)$ to be used in computing entropy as in Eq.~\ref{entropy_truncate}. As in (a) each copy of $\rho(\vec{\theta})$ is prepared by a particular copy of $U(\vec{\theta})$. The measurement results on the ancillary qubit is post-processed for obtaining the said overlap. Note the circuit requires 2 copies of the state to be maintained at a time requiring 2(m+n+1) qubits which retains linear-scaling. After the CSWAP operations, one of the register is freed by resetting the qubits to hold the next copy of the state.}
\label{fig_SWAP_Had}
\end{figure}

It must be emphasized, that computation of the free energy can be efficiently done on a quantum device. For computing the first term $E(\rho(\vec{\theta}, \vec{a}, \vec{b}, \vec{W})) = Tr (H_{is}(\vec{a}, \vec{b}, \vec{W}) \rho(\vec{\theta}))$, the circuit description for the Hadamard Test or its variants\cite{mitarai2019methodology} can be employed. The standard Hadamard test is presented in Fig.~\ref{fig_SWAP_Had}(a). 
In the said figure, state of the single ancillary qubit (single qubit gates of which are marked in purple) is measured and the probability distribution of collapsing it in $|0\rangle$ and $|1\rangle$ are separately recorded. The difference of these sampling probabilities can return the $Tr (H_{is}(\vec{a}, \vec{b}, \vec{W}) \rho(\vec{\theta}))$. The method works well for any unitary operator. It must be noted that for our purpose we use the Pauli decomposition of the Hamiltonian obtained from quantizing Eq.5. Since each term within it is both hermitian and unitary, one can compute the average value of the summand that will be real-valued and coincides with the energy $Tr (H_{is}(\vec{a}, \vec{b}, \vec{W}) \rho(\vec{\theta}))$. 

The computation of $S(\rho(\vec{\theta}, \vec{a}, \vec{b}, \vec{W}))$ can be cumbersome. However, in \cite{PhysRevApplied.16.054035}, the authors have used the convex relaxation of the entropy by truncating at the third order term in the Taylor expansion of the $ln (\rho(\vec{\theta}, \vec{a}, \vec{b}, \vec{W}))$ as follows     
      \begin{equation}
          S(\vec{\theta}, \vec{a}, \vec{b}, \vec{W})  \approx 2Tr(\rho(\vec{\theta}, \vec{a}, \vec{b}, \vec{W})^2) \nonumber - \frac{1}{2}Tr(\rho(\vec{\theta}, \vec{a}, \vec{b}, \vec{W})^3) - \frac{3}{2}\, . \label{entropy_truncate}
      \end{equation}
 The authors showed using such a relaxation of the entropy (and hence the corresponding free energy term in Eq.~\ref{eq:free_energy_cost}), the fidelity of $\tilde{\rho}_{th}$ so procured with the target thermal state $\rho_{th}$ is lower bounded by a function determinable in terms of the inverse temperature $\beta$ and the rank of the target (see Theorem I in~\cite{PhysRevApplied.16.054035}). Such a relaxation has been seen to work well for all the cases in this report too. Furthermore, analytical gradients of the free energy cost with respect to $\vec{\theta}$ is expressible in terms of $E(\rho(\vec{\theta}, \vec{a}, \vec{b}, \vec{W}))$, $Tr(\rho(\vec{\theta}, \vec{a}, \vec{b}, \vec{W})^2)$, $Tr(\rho(\vec{\theta}, \vec{a}, \vec{b}, \vec{W})^3)$ and can be directly obtained from the quantum circuit through parameter-shift rule~\cite{https://doi.org/10.48550/arxiv.1905.13311}. The task at hand is therefore to evaluate terms of the kind $Tr(\rho(\vec{\theta}, \vec{a}, \vec{b}, \vec{W})^k)$ through a quantum circuit ($(k=2,3)$ is used in this work). 

 To this end, several proposals exists which modifies the Hadamard test using CSWAP/Fredkin gates~\cite{PhysRevLett.87.167902,https://doi.org/10.48550/arxiv.quant-ph/0105032,patel2016quantum,cincio2018learning}, proposal mandating the presence of an oracle~\cite{li2018quantum} or density matrix exponentiation~\cite{low2019hamiltonian} exists. However, most such algorithms requires the preparation of $k$-copies of the state $\rho(\vec{\theta}, \vec{a}, \vec{b}, \vec{W})$ simultaneously~\cite{Suba__2019,PhysRevApplied.16.054035}. Recently, Yirka $et al$ \cite{yirka2021qubit} have proposed a qubit-efficient version of evaluating such terms which only requires two copies of the state $\rho(\vec{\theta}, \vec{a}, \vec{b}, \vec{W})$ at once by judicious qubit resetting (see Fig.~\ref{fig_SWAP_Had}(b)). Note that the scheme requires one additional ancillary qubit too. Two copies of the state as prepared using $\vec{U}(\vec{\theta})$ are used at first with a SWAP gate between the $(m+n)$ registers of each (the state $\rho(\vec{\theta}, \vec{a}, \vec{b}, \vec{W})$ is encoded therein). The operation of measuring the trace overlap between the two copies of the state is controlled on the configuration of the extra ancillary qubit. While the first register is retained thereafter, the second set of $(m +n +1)$ qubits are reset to prepare the next copy of the state $\rho(\vec{\theta}, \vec{a}, \vec{b}, \vec{W})$. This process is repeated for all the remaining $(k-2)$ copies too. Thus the circuit uses $2(m + n + 1) +1$ qubits accounting for two copies of the state (each using $(m +n +1)$ qubits) and 1 ancillary qubit. Measurement of the ancillary qubit in $x$-basis followed by classical post-processing can directly return $Tr(\rho(\vec{\theta}, \vec{a}, \vec{b}, \vec{W})^k)$. Apart from implementation on quantum hardwares \cite{https://doi.org/10.48550/arxiv.2002.00055,Suba__2019}, it must also be emphasized that several experimental platforms have also been investigated in recent years which attempts to measure such state-overlaps \cite{PhysRevLett.88.217901, islam2015measuring,PhysRevA.87.052330, PhysRevA.98.052334}. However none of these reports extend the formalism to generative training of a machine learning model as has been done here. Furthermore, such plethora of experimental implementations only highlights and consolidates the ease with which our generative training can be used.  

\bibliography{Supp_Info}